# A Simulation Model for the Waterfall Software Development Life Cycle

Youssef Bassil

LACSC – Lebanese Association for Computational Sciences
Registered under No. 957, 2011, Beirut, Lebanon
youssef.bassil@lacsc.org

## ABSTRACT

Software development life cycle or SDLC for short is a methodology for designing, building, and maintaining information and industrial systems. So far, there exist many SDLC models, one of which is the Waterfall model which comprises five phases to be completed sequentially in order to develop a software solution. However, SDLC of software systems has always encountered problems and limitations that resulted in significant budget overruns, late or suspended deliveries, and dissatisfied clients. The major reason for these deficiencies is that project directors are not wisely assigning the required number of workers and resources on the various activities of the SDLC. Consequently, some SDLC phases with insufficient resources may be delayed; while, others with excess resources may be idled, leading to a bottleneck between the arrival and delivery of projects and to a failure in delivering an operational product on time and within budget. This paper proposes a simulation model for the Waterfall development process using the Simphony.NET simulation tool whose role is to assist project managers in determining how to achieve the maximum productivity with the minimum number of expenses, workers, and hours. It helps maximizing the utilization of development processes by keeping all employees and resources busy all the time to keep pace with the arrival of projects and to decrease waste and idle time. As future work, other SDLC models such as spiral and incremental are to be simulated, giving project executives the choice to use a diversity of software development methodologies.

**Keywords**: Software Engineering, SDLC, Waterfall Model, Computer Simulation, Simphony.NET

## 1. INTRODUCTION

The process of building computer software and information systems has been always dictated by different development methodologies. A software development methodology refers to the framework that is used to plan, manage, and control the process of developing an information system [1]. Formally, a software development methodology is known as SDLC short for Software Development Life Cycle and is majorly used in several engineering and industrial fields such as systems engineering, software engineering, mechanical engineering, computer science, computational sciences, and applied engineering [2]. In effect, SDLC has been studied and investigated by many researchers and practitioners all over the world, and numerous models have been proposed, each with its own acknowledged strengths and weaknesses. The Waterfall, spiral, incremental, rational unified process (RUP), rapid application development (RAD), agile software development, and rapid prototyping are few to mention as successful SDLC models. In a way or another, all SDLC models suggested so far share basic properties. They all consist of a sequence of phases or steps that must be followed and completed by system designers and developers in order to attain some results and deliver a final product. For instance, the Waterfall model, one of the earliest SDLC models, comprises five consecutive phases and they are respectively: Business analysis, design, implementation, testing, and maintenance. On the other hand, the incremental model has seven phases and they are respectively: Planning, requirements, analysis, implementation, deployment, testing, and evaluation [3].

Due to the success of the Waterfall model, many software development firms and industrial manufacturers have adopted it as their prime development framework and SDLC to plan, build, and maintain their products [4]. Additionally, these firms went to the extreme by establishing several departments each of which is run by a team of expert people totally responsible for and dedicated to handle a particular phase of the Waterfall model. This includes, for instance, business and requirements analysis department, software engineering department, development and programming department, quality assurance (QA) department, and technical support department.

However, assigning the exact and the appropriate number of resources for each phase of the Waterfall model including people, equipment, processes, time, effort, and budget was a dilemma and confusion for project managers and directors to achieve the maximum productivity with the minimum number of expenses, workers, and hours. In that sense, it is vital to find the optimal number of resources that should be assigned in order to complete a specific task or phase. For instance, project managers need to find out the number of system analysts that should be hired to work on the business analysis phase. They also need to know how many computers are required for the implementation phase, and how many testers should be acquired to cover all possible test cases during the testing phase. In order to answer all these questions, a simulation for the SDLC is needed so as to estimate the appropriate number of resources necessary to fulfill a certain project of a certain scale.

Relatedly, a computer simulation is a computer program that tries to simulate an abstract model of a particular system. In practice, simulations can be employed to discover the behavior, to estimate the outcome, and to analyze the operation of systems [5].

This paper proposes a simulation model to simulate and mimic the Waterfall SDLC development process from the analysis to the maintenance phase using the Simphony.NET computer simulation tool. The model simulates the different stakeholders involved in the Waterfall model which are essential throughout the whole development process. They include the software solution to design and develop; the employees such as designers and programmers; the different Waterfall phases; and the workflow of every Waterfall task. Furthermore, the proposed simulation takes into consideration three different types of software solutions based on their complexity and scale. The simulation also measures the rate of projects arrival, the rate of projects delivery, and the utilization of various resources during every phase and task.

The goal of the proposed simulation is to identify the optimal number of resources needed to keep the company up with the continuous flow of incoming projects using the minimal amount of workers, time, and budget.

## 2. THE WATERFALL SDLC MODEL

The Waterfall SDLC model is a sequential software development process in which progress is regarded as flowing increasingly downwards (similar to a waterfall) through a list of phases that must be executed in order to successfully build a computer software. Originally, the Waterfall model was proposed by Winston W. Royce in 1970 to describe a possible software engineering practice [6]. The Waterfall model defines several consecutive phases that must be completed one after the other and moving to the next phase only when its preceding phase is completely done. For this reason, the Waterfall model is recursive in that each phase can be endlessly repeated until it is perfected. Fig. 1 depicts the different phases of the SDLC Waterfall model.

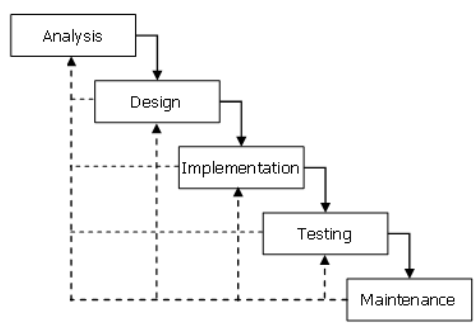

Fig. 1 The Waterfall model

Essentially, the Waterfall model comprises five phases: Analysis, design, implementation, testing, and maintenance.

Analysis Phase: Often known as Software Requirements Specification (SRS) is a complete and comprehensive description of the behavior of the software to be developed. It implicates system and business analysts to define both functional and non-functional requirements. Usually, functional requirements are defined by means of use cases which describe the users' interactions with the software. They include such requirements as purpose, scope, perspective, functions, software attributes, user characteristics, functionalities specifications, interface requirements, and database requirements. In contrast, the non-functional requirements refer to the various criteria, constraints, limitations, and requirements imposed on the design and operation of the software rather than on particular behaviors. It includes such properties as reliability, scalability, testability, availability, maintainability, performance, and quality standards.

Design Phase: It is the process of planning and problem solving for a software solution. It implicates software developers and designers to define the plan for a solution which includes algorithm design, software architecture design, database conceptual schema and logical diagram design, concept design, graphical user interface design, and data structure definition.

Implementation Phase: It refers to the realization of business requirements and design specifications into a concrete executable program, database, website, or software component through programming and deployment. This phase is where the real code is written and compiled into an operational application, and where the database and text files are created. In other words, it is the process of converting the whole requirements and blueprints into a production environment.

Testing Phase: It is also known as verification and validation which is a process for checking that a software solution meets the original requirements and specifications and that it accomplishes its intended purpose. In fact, verification is the process of evaluating software to determine whether the products of a given development phase satisfy the conditions imposed at the start of that phase; while, validation is the process of evaluating software during or at the end of the development process to determine whether it satisfies specified requirements [7]. Moreover, the testing phase is the outlet to perform debugging in which bugs and system glitches are found, corrected, and refined accordingly.

Maintenance Phase: It is the process of modifying a software solution after delivery and deployment to refine output, correct errors, and improve performance and quality. Additional maintenance activities can be performed in this phase including adapting software to its environment, accommodating new user requirements, and increasing software reliability [8].

## 3. RELATED WORK

[9] proposed a simulation planning that must be completed prior to starting any development process. Its purpose is to identify the structure of the project development plan and to classify what must be simulated, the degree of simulation, and how to use the simulation results for future planning. Moreover, the approach takes into consideration such issues as configuration requirements, design constraints, development criteria, problem reporting and resolution, and analysis of input and output data sets. [10] described three types of simulation methodologies. The first is called "simulation as software engineering" and revolves around simulating the delivery of a product. This comprises the use of large simulation models to represent a real system at the production environment. The second is called "simulation as a process of organizational change" and revolves around the delivery of a service. This comprises the use of temporary small-scale models to

simulate small-scale tasks and processes. The third is called "simulation as facilitation" and revolves around understanding and debating about a problem situation. This comprises using "quick-and-dirty" very small-scale models to simulate minute-by-minute processes. [11] proposed the use of simulation as facilitation based on system dynamics. The model proposes the simulation of three development stages: The conceptualization stage which simulates problem situation and system objectives; the development stage which simulates the coding, verification, validation, and calibration processes; and the facilitation stage which simulates group learning around the model, project findings, and project recommendations. [12] proposed a guideline to be followed for performing a simulation study for software development life cycles. It is composed of ten processes, ten phases, and thirteen reliability evaluation stages. Its purpose is to assess the credibility of every stage after simulation and match it with the initial requirements and specifications. The model provides one of the most documented descriptions for simulating life-cycles in the software engineering field [13]. [14] proposed a software engineering process simulation model called SEPS for the dynamic simulation of software development life cycles. It is based on using feedback principles of system dynamics to simulate communications and interactions among the different SDLC phases and activities from a dynamic systems perspective. Basically, SEPS is a planning tool meant to improve the decision-making of managers in controlling the projects outcome in terms of cost, time, and functionalities. [15] proposed a discrete open source event simulation model for simulating the programming and the testing stages of a software development process using MathLab. The model investigates the results of adopting different tactics for coding and testing a new software system. It is oriented toward pair programming in which a programmer writes the code and the simulation acts as an observer which reviews the code and return feedback to the original programmer. In effect, this approach automates the testing and the reviewing processes and promotes best programming practices to deliver the most reliable and accurate code. [16] proposed an intelligent computerized tool for simulating the different phases of a generic SDLC. It is intended to help managers and project directors in better planning, managing, and controlling the development process of medium-scale software projects. The model is based on system dynamics to simulate the dynamic interaction between the different phases of the development process taking into consideration the existence of imprecise parameters that are treated as fuzzy-logic variables.

## 4. PROBLEM DEFINITION & MOTIVATIONS

In practice, software development projects have regularly encountered problems and shortcomings that resulted in noteworthy delays and cost overruns, as well as occasional total failures [17]. In effect, the software development life cycle of software systems has been plagued by budget overrun, late or postponed deliveries, and disappointed customers [18]. A deep investigation about this issue was conducted by the Standish Group [19], it showed that many projects do not deliver on-time, do not deliver on budget, and do not deliver as expected or required. The major reason for this is that project managers are not intelligently assigning the required number of employees and resources on the various activities of the SDLC. For this reason, some SDLC phases may be delayed due to the insufficient number of workers; while, other dependent phases may stay idle, doing nothing, but waiting for other phases to get completed. Consequently, this produces a bottleneck between the arrival and delivery of projects which leads to a failure in delivering a functional product on time, within budget, and to an agreed level of quality.

The proposed simulation for the Waterfall model is aimed at finding the trade-offs of cost, schedule, and functionality for the benefit of the project outcome. It helps maximizing the utilization of development processes by keeping all employees and resources busy all the time to keep pace with the incoming projects and reduce waste and idle time. As a result, the optimal productivity is reached with the least possible number of employees and resources, delivering projects within the right schedule, budget, and conforming to the initial business needs and requirements.

## 5. THE SIMULATION MODEL

This paper proposes a simulation model to simulate the different phases of the Waterfall SDLC model including all related resources, input, workflow, and output. The simulation process is carried out using a simulation tool called Simphony.NET [20] which provides an adequate environment to create, manage, and control the different simulation entities. The purpose of this simulation is to guarantee that the interval-time between each project arrival is equal to the interval-time between each project production. In other words, if a new project is emerging every 10 days, a project must be delivered every other 10 days, taking into consideration that the optimal number of employees should be assigned to every project, that is the number of idle and busy resources should be kept as minimum as possible.

Generally speaking, the proposed simulation process consists of the following steps:
1. Run the simulation, examine the data produced by the simulation,
2. Find changes to be made to the model based on the analysis of data produced by the simulation,
3. Repeat as much as it takes to reach the optimal results.

Technically speaking, the simulation process of the Waterfall model consists of the following steps:
1. Divide the Waterfall model into independent phases,
2. Understand the concept and the requirements that lie behind every phase,
3. Define the resources, tasks, entities, and the work flow of every phase,
4. Simulate each phase apart and record results,
5. Integrate the whole phases together, simulate the system, and record results.

### 5.1. Assumptions and Specifications

Prior to simulating the Waterfall model, a number of assumptions and specifications must be clearly made.

Basically, projects arrive randomly at a software firm with inter-arrival time from a Triangular distribution with a lower limit of 30 days, an upper limit of 40 days, and a mode of 35 days. The probability density function is then given as:

$$f(x|a,b,c) = \begin{cases} 0 & \text{for } x < 30 \\ \frac{2(x-a)}{(b-a)(c-a)} & \text{for } 30 \leq x < 35 \\ \frac{2}{b-a} & \text{for } x = 35 \\ \frac{2(b-x)}{(b-a)(b-c)} & \text{for } 35 < x \leq 40 \\ 0 & \text{for } 40 < x \end{cases}$$

Projects can be divided into three groups based on their complexity and scale: 70% of the projects are small-scale projects, 25% are medium-scale projects, and 5% are large-scale projects.

Each project will require a different mix of specialists, employees, and resources to be delivered based on the scale of the project:
- Small-scale projects require 1 business analyst, 1 designer, 2 programmers, 2 testers, and 1 maintenance man.
- Medium-scale projects require 2 business analyst, 2 designer, 4 programmers, 6 testers, and 2 maintenance man.
- Large-scale projects require 5 business analyst, 5 designer, 10 programmers, 20 testers, and 5 maintenance man.

Assuming that the resources available at the software firm are the following:
- 5 Business Analyst
- 5 Designers
- 10 Programmers
- 20 Testers
- 5 Maintenance Men

And assuming that there exist the following tasks:
- Business Analysis
- Design
- Implementation
- Testing
- Maintenance

And assuming that the duration for every phase to be completed is defined as follows:

The business analysis phase requires a Uniform distribution with a lower limit of 3 days and an upper limit of 5 days.

$$f(x) = \begin{cases} \frac{1}{b-a} & \text{for } 3 \leq x \leq 5 \\ 0 & \text{for } x < 3 \text{ or } x > 5 \end{cases}$$

The design phase requires a Uniform distribution with a lower limit of 5 days and an upper limit of 10 days.

$$f(x) = \begin{cases} \frac{1}{b-a} & \text{for } 5 \leq x \leq 10 \\ 0 & \text{for } x < 5 \text{ or } x > 10 \end{cases}$$

The implementation phase requires a Uniform distribution with a lower limit of 15 days and an upper limit of 20 days.

$$f(x) = \begin{cases} \frac{1}{b-a} & \text{for } 15 \leq x \leq 20 \\ 0 & \text{for } x < 15 \text{ or } x > 20 \end{cases}$$

The testing phase requires a Uniform distribution with a lower limit of 5 days and an upper limit of 10 days.

$$f(x) = \begin{cases} \frac{1}{b-a} & \text{for } 5 \leq x \leq 10 \\ 0 & \text{for } x < 5 \text{ or } x > 10 \end{cases}$$

The maintenance phase requires a Uniform distribution with a lower limit of 1 day and an upper limit of 3 days.

$$f(x) = \begin{cases} \frac{1}{b-a} & \text{for } 1 \leq x \leq 3 \\ 0 & \text{for } x < 1 \text{ or } x > 3 \end{cases}$$

And assuming that each phase upon completion is subject to the following errors:
- There is a 10% probability that a small-scale project will have an error
- There is a 20% probability that a medium-scale project will have an error
- There is a 30% probability that a large-scale project will have an error

### 5.2. The Simphony Model

The proposed simulation model is built using the Simphony.NET simulation tool [20]. In fact, Simphony.NET consists of a working environment and a foundation library that allow the development of new simulation scenarios in an easy and efficient manner. A project in Simphony.NET is made out of a collection of modeling elements linked to each other by logical relationships.

Essentially, the proposed model consists of a set of resource, queue, task, probability branch, capture, release, and counter modeling elements. The resources are the basic employees and workers assigned to work on the phases of the Waterfall model. Each resource has a FIFO queue which accumulates and stores processing events to be processed later. Fig. 2 depicts the resource modeling elements along with their counts and queues. They are respectively the business analyst, the designer, the programmer, the tester, and the maintenance man.

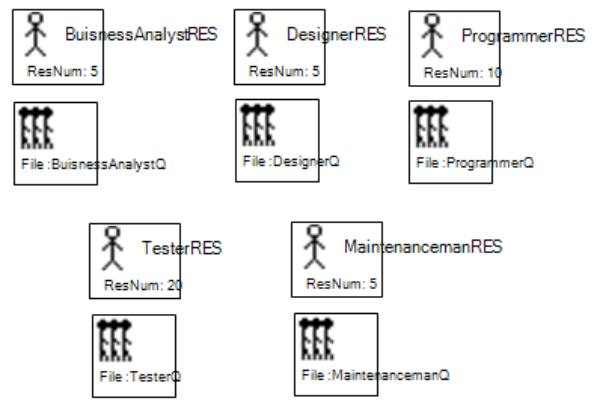

Fig. 2  Resource modeling elements

On the other hand, the Waterfall phases are modeled as a set of task modeling elements each with a capture and release elements. The capture element binds a particular resource to a particular task and the release element releases the resource from the task when it is completed.

Additionally, several probability branch elements exist between the different tasks of the model whose purpose is to simulate the error probability that a Waterfall task might exhibit after completion. The probability element has two branches: Branch 1 with Prob=0.1 denotes that 10% of the small-scale projects are subject to errors; and branch 2 with Prob=0.9 denotes that 90% of the small-scale projects will not exhibit errors after the completion of every phase. These branches simulate the recursive property of the waterfall model to loop over the preceding task if an error was found in the current task.

Moreover, another probability branch element exists at the beginning of every project development cycle whose purpose is to simulate the scale of projects under development. It actually has three branches: Branch 1 with Prob=0.7 denotes that 70% of the incoming projects are small-scale; branch 2 with Prob=0.25 denotes that 25% of the incoming projects are medium-scale; and branch 3 with Prob=0.05 denotes that 5% of the incoming projects are large-scale.

The model starts with a new entity element which sets the number of incoming projects and a counter that counts the number of projects being received, and ends with another counter that counts the number of projects being delivered. Fig. 3 shows the simulation model for the different phases of the Waterfall development process without going deeply into modeling every type of projects. However, Fig. 4 shows the different modeling elements for simulating small-scale type projects.

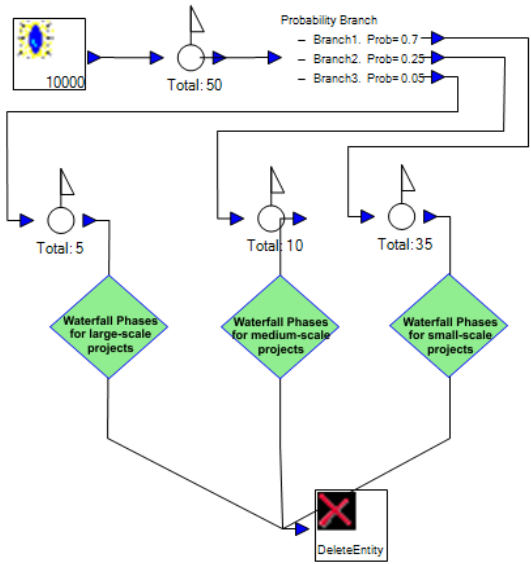

Fig. 3 Simulation model for the Waterfall SDLC

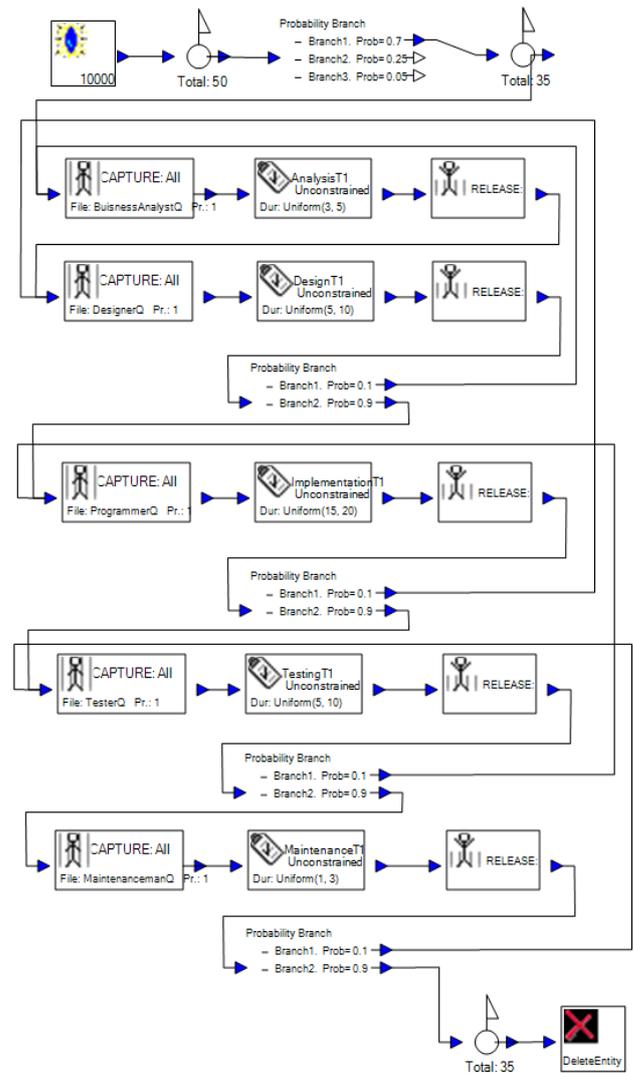

Fig. 4 Simulation model for small-scale type projects

### 5.3. Running the Simulation

The simulation model was executed 5 times, for 1500 milliseconds (2.5 minutes) with 50 incoming projects using the Simphony.NET environment. Table 1 delineates the obtained statistics including the number of projects received and delivered, in addition to the ArT mean time. Table 2 delineates the average utilization of every resource after the completion of the simulation. Furthermore, a graphical representation for resource utilization is plotted in Fig. 5 for the programmer resource; while, Fig. 6 is for the designer resource.

TABLE I
STATISTICS OBTAINED FOR SIMULATING THE WATERFALL MODEL

| small-scale projects received | ArT Mean |
|---|---|
| 35 | 52.09 |
| medium-scale projects received | ArT Mean |
| 10 | 130.45 |
| large-scale projects received | ArT Mean |
| 5 | 426.29 |
| **Total number of projects received: 50** | |
| **Average ArT Mean: 34.46** | |
| small-scale projects delivered | ArT Mean |
| 35 | 53.37 |
| medium-scale projects delivered | ArT Mean |
| 10 | 134.84 |
| large-scale projects delivered | ArT Mean |
| 5 | 448.23 |

| Total number of projects delivered: 50 |
| --- |
| Average ArT Mean: 35.55 |

TABLE II
SIMULATED RESOURCES WITH THEIR AVERAGE UTILIZATION

| Resource | Average Utilization |
| --- | --- |
| Business Analysts | 5.2 |
| Designers | 11.6 |
| Programmers | 21.02 |
| Testers | 7.4 |
| Maintenance Men | 2.09 |

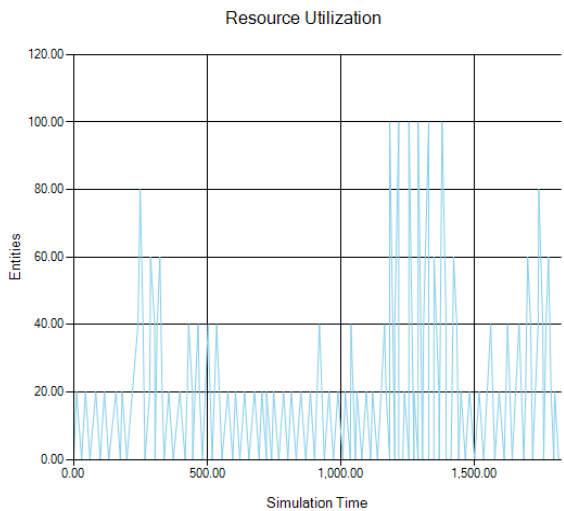

Fig. 5  Utilization of the programmer resource

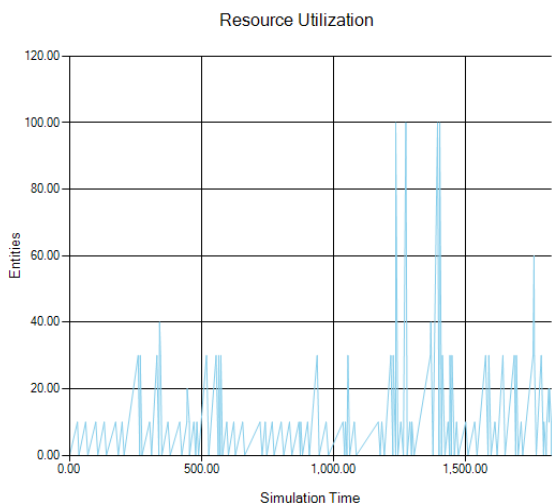

Fig. 6  Utilization of the designer resource

### 5.4. Results Interpretation

The results obtained after running the simulation for many times using the Simphony.NET simulator, clearly showed that the system reached the optimal state when the total number of projects received was equal to the total number of project delivered. In fact, 50 projects were delivered out of 50 without any loss in time or schedule. Additionally, the results helped in pin pointing the optimal number of resources needed to handle the different phases of the waterfall model. The optimal number of required analysts is 5.2, the optimal number of required designers is 11.6, the optimal number of required programmers is 21.02, the optimal number of required testers is 7.4, and the optimal number of required maintenance men is 2.09. These numbers of resources are considered to be the necessary number of workers needed to keep the company up with the continuous flow of incoming projects, in this particular case, dispatching and producing exactly 50 projects on time and within budget.

## 6. CONCLUSIONS & FUTURE WORK

This paper proposed a simulation model for simulating the Waterfall software development life cycle using the Simphony.NET simulator tool. It consists of simulating all entities of the Waterfall model including, software solutions to be developed, operational resources, employees, tasks, and phases. Its aim was to assist project managers in determining the optimal number of resources required to produce a particular project within the allotted schedule and budget. Experiments showed that the proposed model proved to be accurate as it accurately calculated the number of optimal resources required to accomplish a particular software solution based on their utilization metric.

As future work, other SDLC models such as spiral and incremental are to be simulated, allowing project managers to select among a diversity of software development methodologies to support their decision-making and planning needs.

## ACKNOWLEDGMENT

This research was funded by the Lebanese Association for Computational Sciences (LACSC), Beirut, Lebanon, under the "Simulation & Testing Research Project – STRP2012".